# Re-opening open-source science through AI assisted development

Ling-Hong Hung (1), and Ka Yee Yeung (1)
((1) School of Engineering and Technology, University of Washington Tacoma)

**Open-source scientific software is effectively closed to modification by its complexity. With recent advances in technology, an agentic AI team led by a single human can now rapidly and robustly modify large codebases and re-open science to the community which can review and vet the AI generated code.**

## AI agents in software development

Large Language Models (LLMs) began demonstrating promise for software development with vendors such as OpenAI and Anthropic providing increasingly complex models. The arrival of AI agents that can autonomously perform actions, made it possible for AI to execute the **entire software developmental cycle** of planning, coding and testing[1]. While completely autonomous code generation—often colloquially termed 'vibe coding'— can achieve impressive results, it is **not suitable for robust modifications** to large codebases. A major limitation is the **small amount of context** that can be included in a LLM query, causing AI agents to "forget" the details necessary for accurate code generation in complicated, intertwined software[2]. Furthermore, due to the nature of their training sets, AI generated code tends to be brittle, over-engineered, and poorly scalable, and in the worst cases based on hallucinations[3].

## Strategies for improving AI generated code

As developers became familiar with these limitations, strategies evolved to improve AI generated code and make it more reliable.

- **Decomposition** of the problem into smaller modules avoids the agent forgetting details[4]. The modules can be thoroughly tested and then integrated into the codebase.
- **Planning** creates a reference document that can act as long-term memory for more complicated tasks [5]. It allows for vetting of the approach to ensure that the agent is following the user's intent.
- **Testing** is essential for AI generated code. **Unit tests** ensure that the code meets basic design specifications. **Integration tests** ensure that the small decomposed modules still perform after it has been wired into the main application. **Regression tests** ensure that the modifications have not introduced bugs. For agents, this is especially important as they will occasionally alter code that is not directly related to the query.
- **Review** of the generated code correctness, coverage and test results [6]. Agents will still omit details and skip steps in the most clear and comprehensive plan. They will indicate that tests have been passed that have failed. AI code review, by both the human and other agents, is needed to ensure that contracted coding and testing tasks have been properly performed [7].

## Improvements in AI coding models and tooling

Both AI coding models and tools have greatly improved in the latter half of 2025. Context sizes and context management have expanded, allowing models to carry out longer tasks and answer more complicated questions. Hallucination has been dramatically reduced, and the models follow users' instructions much more closely. The frontier reasoning models—OpenAI's GPT-5.1-codex-max and Anthropic's Opus 4.5—have improved to the point that they can provide effective code review and planning in real-time.

Tooling has also advanced. The **Model Context Protocol (MCP)**, introduced by Anthropic at the end of 2024, has become a standard for agent communication[8]. MCP tools for planning, memorializing conversations, querying the internet, and ingesting user files mitigate context limitations and hallucination risks. Other tools, such as coding IDEs, have also improved; Cursor, for example, now provides separate code review and planning tools for AI models and supports multiple simultaneous agents. These improvements, combined with effective AI coding strategies, have made AI agents **viable for serious software development**.

## The Human in the loop - leading a team of AI agents to understand and modify scientific code

The final piece of the puzzle for scientific development is the human in the loop to guide and vet the developmental process. We propose a human-led AI agent team[9] using a carefully staged software development cycle (Figure 1). The **human is the lead Architect** who designs the features, breaks the project into manageable, testable modules, creates gold standard test sets and devises the strategy to assemble the final product. **Thinking Agents** use slow, expensive reasoning models (e.g., Open AI GPT-5-codex-max, Anthropic Opus 4.5) to help the human understand the codebase, identify integration points, and plan and review code. **Coding Agents** use fast cheaper models (e.g., Cursor Composer 1.0, Anthropic Sonnet 4.5) to perform the actual coding and testing. The decomposition, planning, testing and review strategies are used to iteratively build and integrate the changes in a robust manner.

## A real-world example; NIH funded MorPhiC consortium

We present a real-world case study using our proposed AI-assisted development workflow as a member of the NIH funded MorPhiC (Molecular Phenotypes of Null Alleles in Cells) program[10] that aims to advance the understanding of human genes. Our team is tasked with reproducibly processing a variety of data types generated by four data production centers. In this STAR-Flex project, we expanded a STAR codebase from 248 files and 27,785 lines of C++ code to 306 files and 43,849 lines (excluding third party libraries). Flex is a 10x Genomics assay for scRNA-seq that is an inexpensive alternative to traditional single-cell RNA-seq[11]. The use of probes for transcript detection and additional sample barcodes for multiplexing are substantial changes that cannot be handled by the well-established STAR alignment software alone[12]. The only available software to process this data was the proprietary Cell Ranger analysis suite provided by the vendor.

As a public NIH consortium, we aim to **avoid the restrictive EULA** associated with Cell Ranger and develop our own set of open-source modules to ensure transparency and facilitate community contributions. The size of the dataset (2 billion reads) dictated that the project was largely written in C++ for scalability with some Python and R code and orchestrated with STAR using our containerized Biodepot-workflow-builder platform[13].

Our custom pipeline produced large temporary files to handle probe resolution and sample assignment and required some minor modifications of the STAR code. We therefore decided to fully integrate the changes into the STAR binary using the human led AI team strategy described in Figure 1. **The complete refactor was performed by a single scientist with AI assistance, over a period of 6 weeks**, with the final re-integration of the code into a clean STAR fork executed by the AI agents in a single day**.** The final result is a STAR binary with the **full original functionality** and new flags to enable processing of Flex data without the need for external glue code and libraries. The 16,064 new lines of C++ code are **modular, documented, open source** and available for modification and vetting by the community.

**Implications: The Re-opening of Open Source**

We acknowledge that the proposed methodology is not for software novices and requires deep domain knowledge. However, the improvements in AI agentic technology and software developmental workflows are very real and have become powerful tools in the hands of expert developers. A carefully staged and tested AI developmental workflow, when led by an individual scientist, **re-opens open-source** software that had been effectively closed by complexity. The process also enables the community to synergistically vet deficiencies in the code, AI-augmented development restores the original promise of open source: not just transparency, but **modifiability** which is essential for true collaboration. Direct modification of complex codebases is no longer limited to large commercial software teams - AI allows individual researchers to **reclaim control** over their analytical tools. We are no longer limited to indirect changes through orchestration but are able to return to collaborating at the level of the underlying code. While we have demonstrated this using bioinformatics, these conclusions apply broadly to all scientific software. By lowering the technical barrier to entry, AI empowers a broader community to drive rapid software innovation, in response to the next generation of experimental technologies.

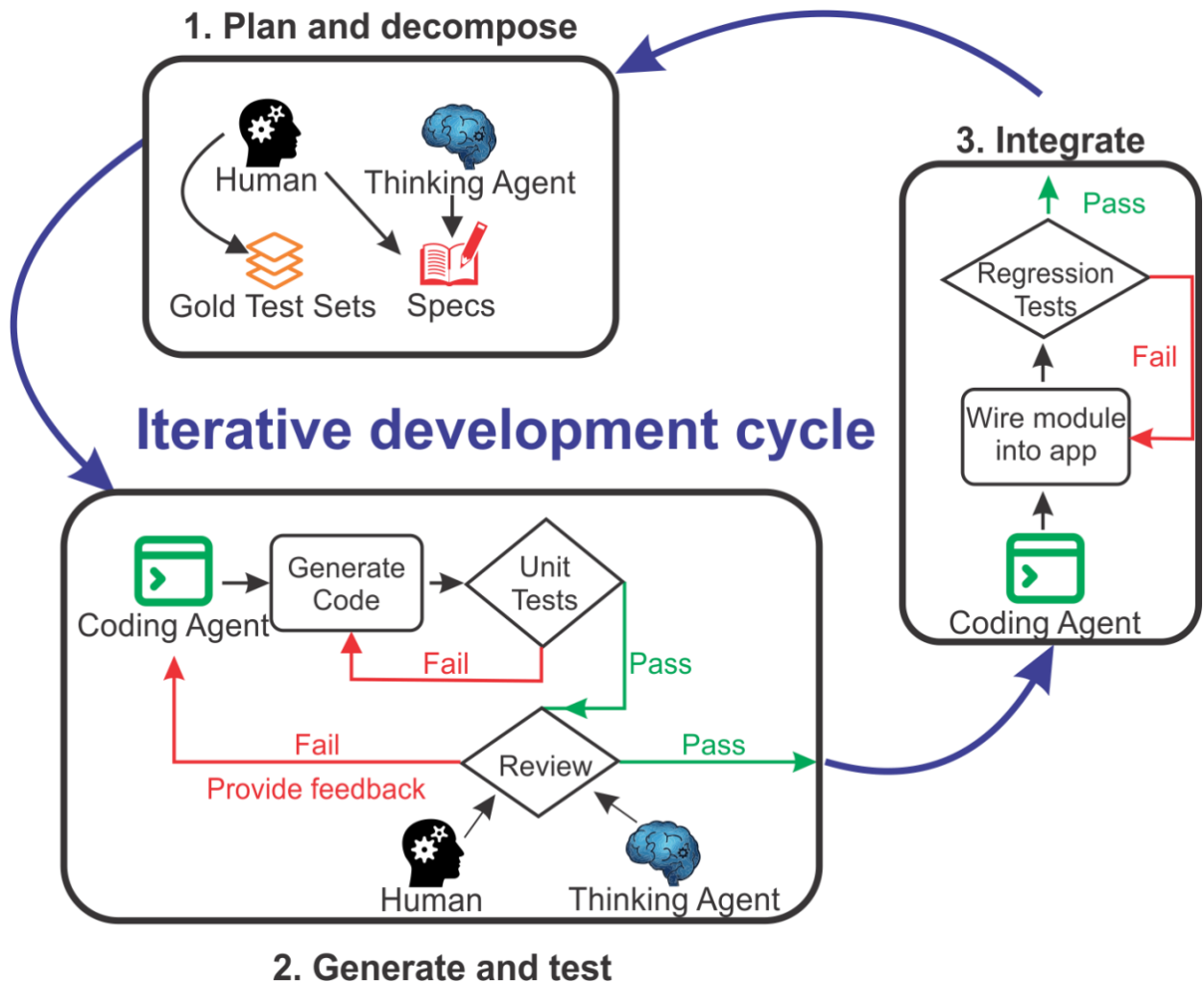

**Figure 1 A human-led, multi-agent software development workflow.** The workflow iteratively adds features in 3 phases: **(1) Plan and Decompose:** The Human Architect and Thinking Agent decompose the scientific goal into atomic modules and define "Gold Test Sets" for validation. **(2) Generate and Test:** The Coding Agent generates code for isolated modules based on technical runbooks and executes unit tests in a clean environment which are evaluated by the Human Architect and the Thinking Agent. **(3) Integrate** Validated modules are reviewed and merged into the legacy core followed by full regression testing by the Coding Agents. Upon successful integration, a final human review is conducted and if necessary, tested against production data for scalability and edge cases. A decision is made to continue refining the implementation or to proceed to the finalization step where unused code is removed and the agents' progress reports are consolidated into final technical documentation.

### Data and code availability

https://github.com/morphic-bio/STAR-Flex

## Funding

L.H.H. and K.Y.Y. are supported by National Institutes of Health (NIH) grant U24HG012674. K.Y.Y. is also supported by the Virginia and Prentice Bloedel Endowment at the University of Washington.